\documentclass[trackchanges, twocolumn]{aastex701}
\usepackage{lmodern}
\usepackage{amsmath}
\usepackage{nccmath}
\usepackage{adjustbox}
\usepackage{soul}
\usepackage{xcolor}
\usepackage{hyperref}

\begin{document}

\title{Infrared Echoes of Precessing Tidal Disruption Events}

%\correspondingauthor{Shao-Yu Fu, Wei-Hua Lei}
%\email{syfu@hust. edu. cn, leiwh@hust. edu. cn}

\author[orcid=0009-0009-5012-9666, sname='Wu']{Hong-Zhou Wu}
\affiliation{Department of Astronomy, School of Physics, Huazhong University of Science and Technology, Luoyu Road 1037, Wuhan, 430074, China}
\email{wuhongzhou@hust.edu.cn}  

\author[orcid=0009-0002-7730-3985]{Shao-Yu Fu}
\affiliation{Department of Astronomy, School of Physics, Huazhong University of Science and Technology, Luoyu Road 1037, Wuhan, 430074, China}
\email[show]{syfu@hust.edu.cn}

\author[orcid=0009-0003-9792-9325]{Wen-Long Xu}
\affiliation{Department of Astronomy, School of Physics, Huazhong University of Science and Technology, Luoyu Road 1037, Wuhan, 430074, China}
\email{D202080113@hust.edu.cn}

\author[0009-0005-9790-1263]{Chang Zhou}
\affiliation{Department of Astronomy, School of Physics, Huazhong University of Science and Technology, Luoyu Road 1037, Wuhan, 430074, China}
\email{czhou@hust.edu.cn}

\author[orcid=0000-0003-3440-1526]{Wei-Hua Lei}
\affiliation{Department of Astronomy, School of Physics, Huazhong University of Science and Technology, Luoyu Road 1037, Wuhan, 430074, China}
\email{leiwh@hust.edu.cn}

\begin{abstract}
%Tidal disruption events (TDEs) represent powerful cosmic laboratories for studying supermassive black holes and their circumnuclear environments. 
A tidal disruption event (TDE) occurs when a star is torn apart by a supermassive black hole. The resulting UV/optical flare irradiates parsec-scale dust, producing delayed mid-infrared echoes that persist for years. These echoes provide unique calorimetric probes of the total radiated energy and dust geometry.
%, largely independent of orientation biases that plague X-ray and UV observations.  predominantly
Existing models usually assume static axisymmetric illumination patterns. However, the TDE accretion disk is likely misaligned and undergoes relativistic precession.
%However, existing models usually assume static, axisymmetric illumination patterns, neglecting the rich phenomenology expected from misaligned accretion disks that undergo relativistic precession. 
In this work, we present a theoretical framework for infrared dust echoes from a precessing TDE disk. The precession will lead to highly variable infrared light curves, which can be revealed by high-cadence observations. The overall profile of the infrared light curves shows double-peaked to single-peaked pattern transitions as a result of the changes in the viewing angle or precession angle.
%We find that both the viewing angle and the precession angle can lead to the transition from double-peaked to single-peaked echo light curves. The high-cadence infrared observations will reveal the 
%We find that the viewing angle controls the transition from double-peaked to single-peaked echo morphology as the observer approaches edge-on. Crucially, at intermediate viewing angles, precession superimposes transient plateau-like features between the two peaks—arising from epochs where the time-varying illuminated region temporarily approximates ring-like geometry. The coexistence of a broad, flattened peak and embedded rapid, high-amplitude variability in the light curve provides a distinctive signature of precession in TDE dust echoes.
The results indicate that infrared echoes are dynamic tracers of the evolving lighting patterns of the central engine. 
%Our model may provide some assistance for the work of JWST and the Rome Space Telescope. 

\end{abstract}


\section{INTRODUCTION}
When a star ventures sufficiently close to a supermassive black hole (SMBH) residing in a galactic nucleus, the tidal forces exerted by the SMBH can overwhelm the star's self-gravity, leading to its disruption. Such tidal disruption events (TDEs) liberate a tremendous amount of gravitational binding energy, with typical UV-optical flares reaching peak luminosities of $10^{44}-10^{45} \, \text{erg s}^{-1}$ and total energies of $10^{51}-10^{52} \, \text{erg}$ on timescales of months to years \citep{Rees1988, Phinney1989, Lu2016, Gezari2021}. The resulting accretion of stellar debris onto the SMBH produces luminous, multi-wavelength transients that serve as powerful tools for probing otherwise quiescent black holes and the circumnuclear environments of their host galaxies. The observational signatures of TDEs are diverse, spanning soft X-ray, ultraviolet (UV), optical, and increasingly, infrared (IR) wavelengths \citep{guolo2024systematicanalysisxrayemission, Lin_2024}. 

The central parsecs of galaxies are permeated by distributions of circumnuclear dust, residing in structures ranging from toroidal obscurers to more diffuse, filamentary configurations \citep{Alexander2020, Mou2022, Bu2023, Lei2024, Zhuang2025, Zhou2026, An2026}. When a sudden UV-optical flash---such as that from a TDE---illuminates this dust, the dust grains absorb the high-energy photons and re-radiate the energy thermally in the mid-infrared (mid-IR), typically peaking at wavelengths of $3-10 \, \mu\text{m}$. This phenomenon, termed a dust echo, manifests as a delayed, extended infrared transient that can persist for years after the primary flare has faded \citep{Lu2016, van_Velzen_2016}. Typical delays of months to years correspond to dust located at parsec scales \citep{van_Velzen_2016}. The IR light curve shape, luminosity, and spectral energy distribution (SED) encode critical information about the dust geometry, temperature distribution, and covering factor. For TDEs, dust echoes have been observed to reach luminosities of $10^{42}-10^{43} \, \text{erg s}^{-1}$, with dust temperatures spanning $900-2500 \, \text{K}$ and covering factors of order tens of percent \citep{Jiang_2017, Dou_2017}. The temporal evolution often follows an initial linear rise---reflecting dust of increasing distance on the near side of the system---followed by a plateau or gradual decline as the echo fades. 

The causal connection between TDEs and their associated mid-IR echoes is both direct and physically robust. In this framework, the TDE provides the initial UV-optical flash that irradiates surrounding dust, while the spatial and structural properties of the circumnuclear dust distribution govern the morphology and temporal evolution of the resulting echo. This physical relationship has been clearly established in multiple studies: the TDE emits a luminous flare across UV and optical wavelengths, which illuminates dust located at parsec-scale distances; the dust grains absorb this radiation and re-emit the energy thermally in the mid-IR, enabling constraints on key dust properties such as the covering factor, temperature profile, and potential signatures of silicate or polycyclic aromatic hydrocarbon (PAH) features---particularly in weakly active galactic nuclei \citep{Lu2016}. The dust covering factor, denoted as $f_{\text{dust}}$, typically estimated to be $\sim 10\%$ for TDEs, can be derived from the ratio of the total re-radiated infrared energy to the incident UV/optical energy input \citep{Jiang_2017, van_Velzen_2016}. This calorimetric nature renders IR echoes uniquely valuable as probes of the total radiative energy budget, offering measurements that are largely independent of orientation-dependent biases that commonly affect direct X-ray or UV observations. 

Observational support for this model has grown significantly in recent years. The first confirmed mid-IR dust echo was identified in the ultra-luminous infrared galaxy (ULIRG) F01004-2237, where a candidate TDE exhibited a delayed infrared response consistent with dust reprocessing \citep{Dou_2017}. Follow-up analyses of additional TDE candidates, such as PS16dtm, have yielded more precise constraints: an IR flare peaking near $\sim 3 \, \mu\text{m}$, with a time delay of approximately 11 days relative to the optical peak, implying a compact dust distribution within a radius of less than 10 light-days and dust temperatures ranging from $900$ to $2500 \, \text{K}$ \citep{Jiang_2017}. Furthermore, long-term monitoring of transient coronal line emitters has revealed that IR emission can persist for several years, often exhibiting a plateau phase followed by a gradual decline, consistent with extended dust distributions being progressively illuminated and cooled \citep{van_Velzen_2021}. Systematic surveys utilizing WISE data have now uncovered dozens of candidate dust echoes linked to TDEs and changing-look active galactic nuclei (AGN), firmly establishing mid-IR echoes as a powerful diagnostic tool for identifying obscured nuclear transients and mapping the structure of circumnuclear environments \citep{Dou_2017, Necker_2025, Reynolds_2022}. 

Parallel to these advances, a new class of nuclear transients---quasi-periodic eruptions (QPEs)--has emerged, characterized by high-amplitude, nearly periodic X-ray and UV flares recurring on timescales of hours to days \citep{Shu2013, Miniutti2019, Giustini2020, 2021Natur.592..704A, Chakraboty2021, Arcodia2024, Nicholl2024, Hernandez2025}. First discovered by \citet{2019Natur.573..381M} in GSN 069 and subsequently studied by \citet{2021Natur.592..704A}, QPEs are interpreted as repeated partial disruptions of a star on an eccentric orbit or instabilities in a remnant accretion disk \citep{Shu2018}. Critically, a recent seminal study predicted that QPEs should generate detectable IR echoes, with time delays of order one year and distinct observational signatures that can differentiate between QPE models \citep{pasham2025usinginfrareddustechoes}. This prediction opens a new frontier: using IR echoes not only as passive probes of dust, but as active diagnostics of the central engine's geometry and dynamics. 

Despite these successes, existing dust echo models predominantly assume isotropic or axisymmetric dust distributions and non-precessing, single-impulse illumination \citep{pasham2025usinginfrareddustechoes, van_Velzen_2021, Winter_2023}. While adequate for many TDEs, these simplifications may break down for events with complex central engine geometries or recurring outbursts. The precession of an accretion disk or outflow---driven by relativistic frame-dragging (Lense-Thirring precession) or torques from a misaligned binary---can systematically alter the direction of the illuminating radiation pattern \citep{LenseThirring1918, 2007ApJ...668..417F, 2012PhRvL.108f1302S, Lei2013, Teboul2023, Wang2025SciA, Lu2024, Li2025, Chen2026, Yuan2026}. This, in turn, should imprint a distinctive, time-dependent anisotropy on the IR echo, modulating both its intensity and spectral evolution in ways unaccounted for in static models.

While direct imaging of TDE disk precession remains challenging, growing multi-wavelength observational evidence suggests that misaligned, precessing disks may be common and imprint periodic signatures across the electromagnetic spectrum. In the X-ray and radio bands, AT2020afhd exhibits 19.6-day synchronized X-ray and radio variations interpreted as disk-jet co-precession driven by Lense-Thirring torques \citep{Wang2025SciA}. In the optical regime, AT2019aalc displays quasi-periodic oscillations with periods of $\sim$ 162 days and $\sim$ 88 days across ZTF bands, accompanied by quasi-periodic swings of the linear polarization angle ($\Delta \theta \approx 40^{\circ}$) phase-locked to the flux maxima---both signatures attributed to a precessing inner disk \citep{Jordana_Mitjans_2025}. In the NUV, eRASSt J0456-20 shows periodic variability anti-correlated with X-ray bursts, consistent with a precessing super-Eddington disk \citep{Chen2026}. However, not all TDEs exhibit optical periodicity even when the  disk is precessing. AT2020ocn, for instance, shows $\sim$ 15-day X-ray QPOs attributed to Lense-Thirring precession, yet its optical/UV emission remains aperiodic---likely because it originates from an optically thick, quasi-spherical reprocessing envelope that washes out the central engine's geometric modulation \citep{Pasham_2024}. Infrared dust echoes, by contrast, respond directly to the anisotropic UV/optical illumination from the accretion disk, making them uniquely sensitive to precession-induced anisotropies on year-long timescales.

The inspiration for such a model comes directly from the QPE paradigm: if QPEs, with their periodic central engine activity, can produce observable IR echoes, then a precessing TDE---where the effective illumination direction slowly slews across the sky---should yield an even richer echo signature. Indeed, the slow precession of a TDE accretion disk, potentially extending over years, introduces a geometric modulation of the dust illumination that is qualitatively distinct from both single-flare TDEs and periodic QPEs. By modeling this effect, we can potentially constrain the precession period, the dusty torus geometry, and the anisotropy of the TDE flare itself. 

We studied how precession qualitatively alters TDE infrared echoes. In  \autoref{sec:2}, we discussed different dust models under non precession scenarios. In  \autoref{sec:3}, we introduced a simplified precession model with a constant angular velocity and some of our results. In \autoref{sec:4}, we made predictions. 
\section{MODELS} \label{sec:2}

In this section, we present our model of circumnuclear dust echoes in TDEs. We restrict the model to absorption and re-emission processes, not including scattering. This is justified because, (1) given that the spectral peak of the TDE flare lies at UV/optical wavelengths, the contribution of the scattered TDE light at mid-IR wavelengths can be negligible, compared to that of thermal re-emission from circumnuclear dust, and (2) in the distribution of dust of interest in this paper, further scattering of the re-emitted IR photons is negligible. To address dust temperature evolution, we adopt the assumption of instantaneous thermal equilibrium. This is well justified because the heat capacity of a single dust grain is extremely small, rendering its radiative cooling time negligible on all physically relevant timescales \citep{Gerardy_2002}. While computational methods for an echo from spherically distributed dust \citep{Dwek1983, Dwek1985} and further from aspherically distributed dust \citep{Dwek1989, Felten1989} have been developed, we assume for the sake of simplicity that the distribution of circumnuclear dust is represented by an infinitely thin shell located at radius $R$ from the central black hole. The thin-shell approximation further simplifies the computation as shown below. Furthermore, in providing an upper limit for the amount of circumnuclear dust as a function of $R$, the thin-shell approximation basically provides the most conservative limit, as any spherically symmetric distribution can be described as a convolution of different shells \citep{maeda2015constrainingcircumstellarmatterdust}. Namely, if one adds another shell at $R'$, that is different from $R$ for which the constraint is considered, then this additional shell will increase the predicted flux without contributing to the dust mass at $R$; therefore, the upper limit on this dust mass should be reduced (i.e., the predicted flux must be below observations). 
Below, we present the general expression for infrared echoes under arbitrary observation directions. 

Let us first consider a scenario where, in a standard spherical coordinate system, we observe a dust shell with a temperature of $T (\theta) $ and a mass of $M_d$. We then specify the unit vector of the observer's direction 
\begin{align}
    \hat{\boldsymbol{n}}= (sin \theta _{\rm obs} cos\phi_{\rm obs}, sin\theta_{\rm obs}sin\phi_{\rm obs}, cos\theta_{\rm obs}) 
\end{align}
and the direction of the shell's surface element 
\begin{align}
    \hat{\boldsymbol{r}}= (sin \theta  cos \phi , sin\theta sin\phi, cos\theta) 
\end{align}
to obtain the relative delay time
\begin{align}
    t'=t-\frac{R}{c} (1-\hat{\boldsymbol{n}} \cdot \hat{\boldsymbol{r}}) 
\end{align}
Then, the luminosity of the thermal emission from dust is given as follows:
\begin{align}
\nonumber
L_{\mathrm{echo}, \nu} & =\int_V 4 \pi \kappa_{\mathrm{a}, \nu} B_\nu (T (\theta) ) \rho d V 
\end{align}
%\\ &=2 \pi M_{\mathrm{d}} \kappa_{\mathrm{a}, \nu} \int_0^\pi \sin \theta B_\nu (T (\theta) ) d \theta

\begin{figure*}[htbp]
    \centering
    \includegraphics[width=\textwidth]{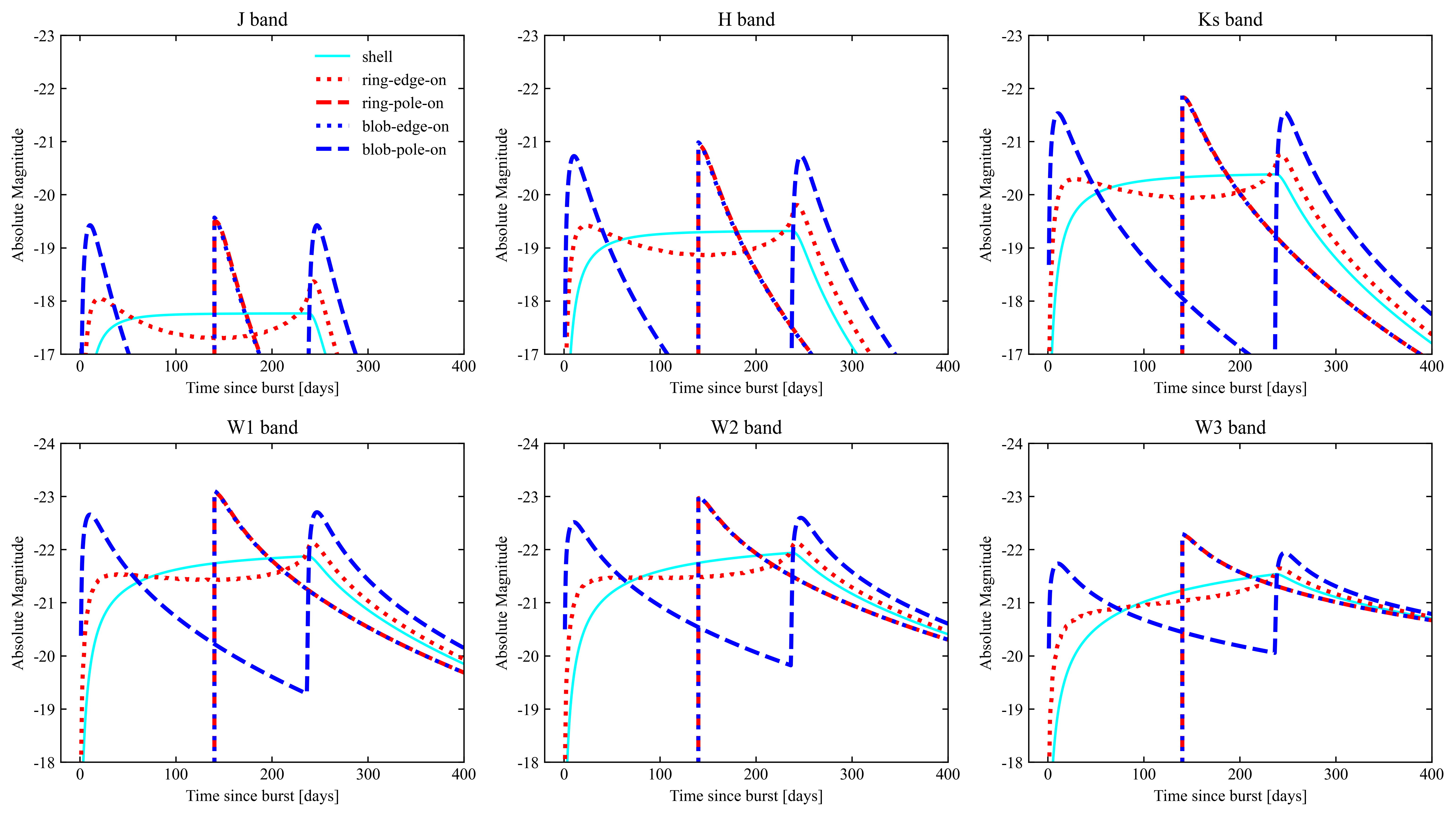}
    \caption{The absolute magnitude of different geometric shapes of dust on the echo light curve. For demonstration purposes, the distance from the dusty torus is fixed at $R$=0.1 pc. The reference shell model is displayed in cyan (solid). The echo light curve from the circular dusty torus distribution is represented by a red curve, which represents the observer at the edge (dotted line) and the observer at the pole (dashed line). The speckled dusty torus distribution represented by the blue curve also applies to observers on the edges (dotted line) and poles (dashed line). }
    \label{fig:output}
\end{figure*}

Here, $\kappa_{a, \nu}$ is the absorbing opacity of the dust at frequency $\nu$. It is specified by the dust properties, which are assumed to be uniform within the shell. Furthermore, the shell is assumed to have uniform density ($\rho$). The above expression can be converted into time-dependent luminosity by introducing $t$ (the time since the explosion, measured in the observer’s frame) and $t'$ (observer-frame time when the light was emitted from a given volume element at $\theta$ as observed at time $t$). 

Combining Eqs.(3) and (4), we obtain the following expression for the echo luminosity at time $t$:
\begin{align}
    \nonumber
    L_{\rm echo, \nu}\left (t ; \theta_{\mathrm{obs}}, \phi_{\mathrm{obs}}\right) =\frac{c}{4 \pi R} M_d \kappa_{a, \nu} \int_0^\pi \sin \theta d \theta \\
    \times \int_0^{2 \pi} d \phi B_\nu\left (T\left (t-\frac{R}{c}[1-\hat{\boldsymbol{n}} \cdot \hat{\boldsymbol{r}}]\right) \right) 
\end{align}
Where $M_{d}=4\pi \Sigma_{d} r^2$, $\Sigma_{d}$ represents the surface density of the accretion disk. 

The temperature evolution of the dusty torus at time $t$, under radiative equilibrium with an incoming TDE flux, $L_{\rm TDE, \nu} (t) $, is given by
\begin{align}
\int_0^\infty \frac{L_{TDE, \nu} (t) }{4 \pi R^2} \kappa_{a, \nu} = 4 \pi \int_0^\infty \kappa_{a, \nu} B_{\nu} T (t) d \nu
\end{align}

\subsection{Shell Model} \label{subsec:shell}
For the Shell model (spherically symmetric dust shell), the echo luminosity is independent of the observation direction under the assumption that the geometry is perfectly spherically symmetric and the temperature depends only on time (not on direction). The relationship between $t$ and $t'$ is given by the following:
\begin{align}
    t'=t-\frac{R}{c} (1-cos\theta ) 
\end{align}
The expression for the echo luminosity at time $t$ degenerates to:
\begin{align}
L_{\mathrm{echo}, \nu}^{shell} (t) =2 \pi \frac{c}{R} M_{d} \kappa_{a, \nu} \int_{max (t- \frac{2R}{c}, 0) }^t B_{\nu} (T (t') ) d t'
\end{align}

\subsection{Ring Model}
\label{subsec:ring}
The dusty torus around a TDE  may be confined within the equatorial plane \citep{Wu_2025}, therefore we also consider a torus/ring-like structure, again under the infinitely thin assumption in the radial direction. We denote $\theta _0$ as the opening angle of the ring, as measured from the equatorial direction. For a fixed $\theta $, the ring is only visible when $\phi$ satisfies condition $cos (\phi-\phi_{\rm obs}) \in [\alpha, 1]$, where
\begin{align}
    \alpha (\theta, t') \equiv
\frac{1 - \frac{c}{R} (t-t') - \cos\theta_{\text{obs}}\cos\theta}{\sin\theta_{\text{obs}}\sin\theta}
\end{align}
The echo luminosity is expressed as follows
\begin{fleqn}[0pt]
\begin{align}
\nonumber
&L_{echo, \nu}^{\text{ring}} (t;\theta_{\text{obs}}) =\frac{c}{2\pi R}\frac{M_d}{\sin\theta_0}\kappa_{a, \nu} \times\\
&\quad\quad\quad\quad\int_{t'_{\min}}^{t'_{\max}}
\int_{\frac{\pi}{2}-\theta_0}^{\frac{\pi}{2}+\theta_0}
\frac{\beta}{\sqrt{1-\alpha^2}}\, 
B_\nu (T (t') ) \, d\theta\, dt' 
\end{align}
\end{fleqn}
where
\begin{gather}
    \nonumber
    t'_{\min} = t - \frac{R}{c}\left (1 + \sin\theta_{\text{obs}}\sin\theta_0 + |\cos\theta_{\text{obs}}|\cos\theta_0\right) \\
    \nonumber
    t'_{\max} = t - \frac{R}{c}\left (1 - \sin\theta_{\text{obs}}\sin\theta_0 - |\cos\theta_{\text{obs}}|\cos\theta_0\right) 
\end{gather}

\begin{gather}   \nonumber\beta=2\cos^{-1}\!\bigl (\max (-1, \min (1, \alpha) ) \bigr) 
\end{gather}

For an observer sitting at the pole-on position, the echo luminosity is expressed as follows:
\begin{fleqn}[0pt]
\begin{align}
\nonumber
&L_{\mathrm{echo}, \nu} (t) =2 \pi \frac{c}{R} \frac{M_{\mathrm{d}}}{\sin \theta_0} \kappa_{\mathrm{a}, \nu} \times \\
&\quad\quad\quad\quad\quad\int_{max (t-\frac{R}{c} (1+sin\theta_0), 0) }^{t-\frac{R}{c} (1-sin\theta_0) } B_{\nu} (T (t') ) dt'
\end{align}
\end{fleqn}

If $t-\frac{R}{c} (1+sin\theta_0) <0$, then the integral is set to zero. Finally, the echo luminosity arising from the same configuration, but viewed edge-on, is given as follows. 
\begin{align}
\nonumber
L_{\mathrm{echo}, \nu} (t) & =2 \pi \frac{c}{R} \frac{M_{\mathrm{d}}}{\sin \theta_0} \kappa_{\mathrm{a}, \nu}\left[\int_{\max \left (t-\frac{2 R}{c}, 0\right) }^{t-\frac{R}{c}\left (1+\cos \theta_0\right) } B_\nu\left (T\left (t^{\prime}\right) \right) d t^{\prime}\right. \\
\nonumber
\quad+ & \int_{t-\frac{R}{c}\left (1+\cos \theta_0\right) }^{t-\frac{R}{c}\left (1-\cos \theta_0\right) } \frac{2}{\pi} B_\nu\left (T\left (t^{\prime}\right) \right) \sin ^{-1}\left (\frac{\sin \theta_0}{\sin \theta}\right) d t^{\prime} \\
\quad+ & \left. \int_{t-\frac{R}{c}\left (1-\cos \theta_0\right) }^t B_\nu\left (T\left (t^{\prime}\right) \right) d t^{\prime}\right]
\end{align}

\begin{figure*}[ht!]
\plotone{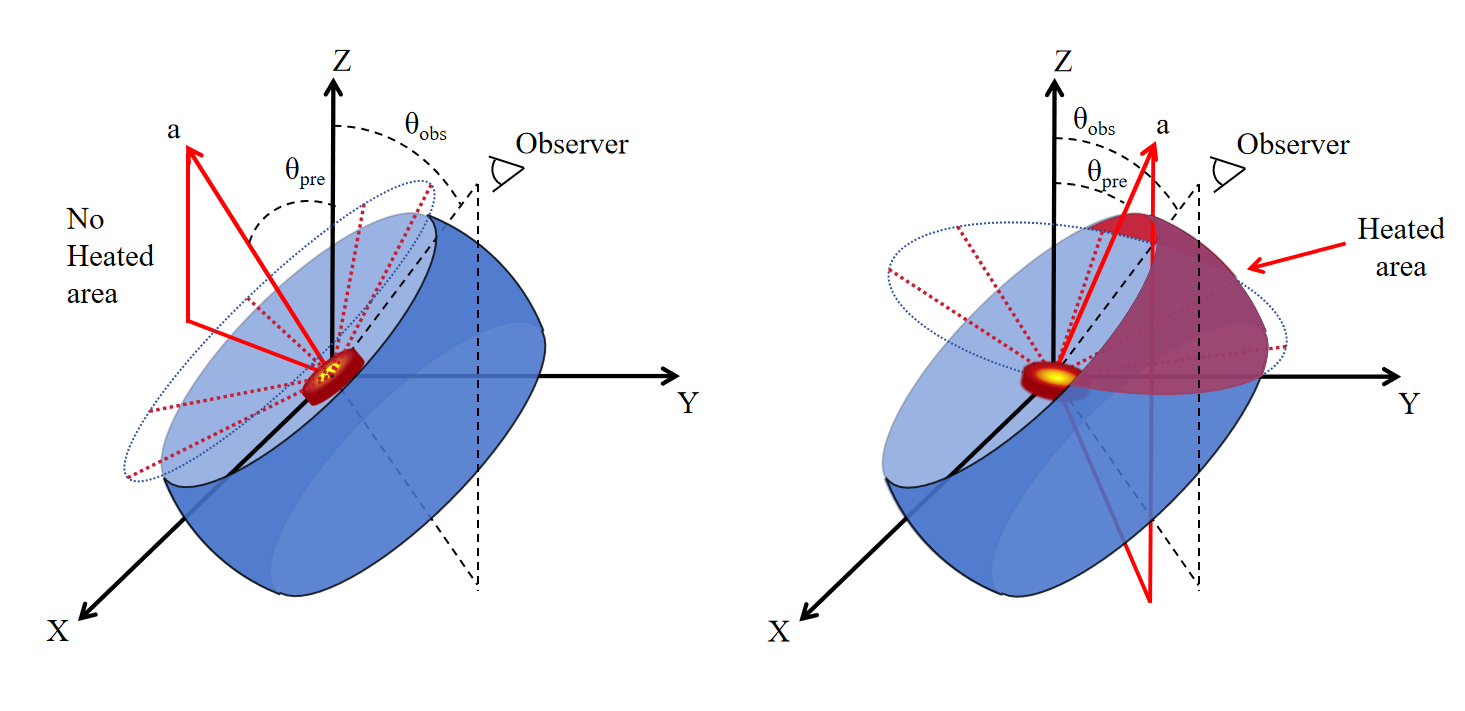}
\caption{Schematic geometry of the precessing TDE dust-echo model. The black-hole spin axis defines the Z-direction. The accretion disk's spin axis (a) is tilted by a constant precession angle $\theta_{\rm pre}$ and precesses around Z with angular frequency $\Omega$, causing the TDE's UV/optical illumination beam to sweep across the circumnuclear dusty torus. The torus has a finite opening angle $\theta_0$  and is approximated as an infinitesimally thin shell at radius $R$. The observer's line of sight makes an angle $\theta_{\rm obs}$ with respect to Z. At any given time $t$, only a localized patch of dust (red "heated area") whose surface normal lies within the disk's instantaneous radiation cone is efficiently heated, producing a blob-like IR echo signature as the precessing beam sweeps the torus. }. 
\label{fig:Model}
\end{figure*}

\subsection{Blob Model}
\label{subsec:blob}
Another interesting possibility for the dusty torus distribution is a bipolar morphology. Therefore, we also consider a pair of bipolar blobs for the distribution of the dusty torus, again under the infinitely thin assumption in the radial direction. We denote $\theta_0$ as the opening angle subtended by each blob, as measured from the polar direction. The echo luminosity is expressed as follows
\begin{align}
\nonumber
&L_{echo, \nu}^{\text{blob}} (t;\theta_{\text{obs}}) =
\frac{c}{4\pi R}\frac{M_d}{1-\cos\theta_0}\kappa_{a, \nu}\times\\&
\int_{t'_{\min}}^{t'_{\max}}
\left[
\int_{0}^{\theta_0} + \int_{\pi-\theta_0}^{\pi}
\right]
\frac{\beta}{\sqrt{1-\alpha^2}}\, 
B_\nu (T (t') ) \, d\theta\, dt'
\end{align}

For an observer sitting at the pole-on position, the echo luminosity is expressed as follows:

\begin{flalign}
& L_{\text{echo}, \nu} (t) = 2\pi \frac{c}{R} \frac{M_{\mathrm{d}}}{1-\cos\theta_0} \kappa_{\mathrm{a}, \nu} 
\Bigg[ \, \int_{\max\left (t-\frac{2R}{c}, 0\right) }^{t-\frac{R}{c} (1+\cos\theta_0) } 
\left (T (t') \right) \, dt' \nonumber \\
& \qquad + \int_{t-\frac{R}{c} (1-\cos\theta_0) }^{t} \left (T (t') \right) \, dt' \, \Bigg] &&
\end{flalign}

The echo luminosity arising from the same configuration, but viewed edge-on, is given as
\begin{align}
\nonumber
L_{\mathrm{echo}, \nu} (t) & =2 \pi \frac{c}{R} \frac{M_{\mathrm{d}}}{1-\cos \theta_0} \kappa_{\mathrm{a}, \nu} \int_{t-\frac{R}{c}\left (1+\sin \theta_0\right) }^{t-\frac{R}{c}\left (1-\sin \theta_0\right) } \\
\quad \times & \frac{2}{\pi} B_\nu\left (T\left (t^{\prime}\right) \right) \cos ^{-1}\left (\frac{\cos \theta_0}{\sin \theta}\right) d t^{\prime}
\end{align}

\subsection{Results} \label{subsec:result}

In order to more simply and intuitively observe the influence of different structures on the infrared echo light variation, we only provide light variation curves at a few specific observation angles. \autoref{fig:output} shows the predicted NIR to MIR light curves (J, H, Ks, W1:3.4$\mu$m, W2:4.6$\mu$m, and W3:12$\mu$m) for fixed radius values $R (0.1 \text{pc}) $ and $M_d (3 M_{\odot}) $ , the effects of different geometries are demonstrated; light curves are shown for the shell dust, pole-on and edge-on views for a ring-like dusty torus with $\theta_0 = 10^{\circ}$, and pole-on and edge-on views for bipolar blobs with $\theta_0 = 10^{\circ}$. 

For a fixed total dust mass, confining the material to a restricted solid angle enhances the local density, producing higher luminosities at specific epochs compared to the spherical shell model and thereby enabling tighter constraints on $M_{\rm d}$ when observations coincide with these phases. The shell model exhibits a viewing-angle-independent light curve with three distinct stages: an initial rise as the illuminated area expands, a plateau upon full illumination of the shell, and a final decline tracing the TDE flare decay. In contrast, geometric anisotropy in the ring and blob models introduces strong viewing-angle dependence. The edge-on ring model produces a plateau with elevated edges---featuring a faster rise and lower plateau luminosity than the shell due to the smaller surface area, plus small peaks at the plateau edges from time-delay effects---while the pole-on view yields a single steep peak with shorter duration but higher maximum brightness achieved when emission from the ring edge arrives. Similarly, the edge-on blob model displays a steep single peak as both clumps arrive nearly simultaneously, whereas the pole-on perspective produces two distinct sharp peaks due to the large path difference between clumps. 

\section{Precessing TDE Model}
\label{sec:3}

Illustrated by the geometric sketch in \autoref{fig:Model}, we present a precessing disk-dust interaction model. We consider a circumnuclear dust distribution morphologically analogous to a toroidal shell (torus), characterized by a narrow radial extent but non-confinement to the equatorial plane. While the dust is assumed infinitesimally thin radially to simplify the computation of equal-delay surfaces, its finite vertical opening angle $\theta_0$---measured from the edge-on direction of the torus---introduces critical geometric flexibility. This parameter $\theta_0$ governs the latitudinal coverage of the dust, bridging the extremes of a razor-thin ring ($\theta_0 \to 0^{\circ}$) and a fully spherical shell ($\theta_0 \to 90^{\circ}$), thus allowing exploration of realistic torus geometries observed in AGN and quiescent galactic nuclei. 

The observer's frame is defined by the line-of-sight angle $\theta_{\rm obs}$ relative to the black-hole spin axis (Z). Crucially, we introduce relativistic precession by tilting the accretion disk's spin axis $\mathbf{a}$ by a constant angle $\theta_{\rm pre}$ with respect to the black-hole spin axis. This misalignment drives the disk to precess about the Z-axis at angular frequency $\Omega_{\rm LT}$ (Lense-Thirring frequency) or $\Omega_{\rm binary}$ (if torqued by a secondary SMBH). At any instant $t$, the heating beam---defined by the disk's instantaneous radiation pattern---sweeps across the dusty torus, illuminating only a localized patch. 

As depicted in \autoref{fig:Model}, the precessing illumination creates a time-dependent hot spot on the dust surface. Due to the highly anisotropic nature of TDE's UV/optical flashes (which preferentially illuminate along the disk normal), only dust elements aligned with the instantaneous disk surface can be effectively heated. This geometric selection effect collapses the echo response into a compact, blob-like signature in the time domain: rather than a broad, axisymmetric echo from a static illumination pattern, the observed IR light curve traces the convolution of a moving heat source with the torus geometry, producing sharp, asymmetric peaks whose arrival time, width, and amplitude modulate with the precession period. This blob-like morphology is the smoking-gun signature of precession, distinguishing it from static or isotropic TDE models where the echo would instead exhibit broad, symmetric plateaus arising from full-surface illumination.

\begin{figure*}[ht!]
\plotone{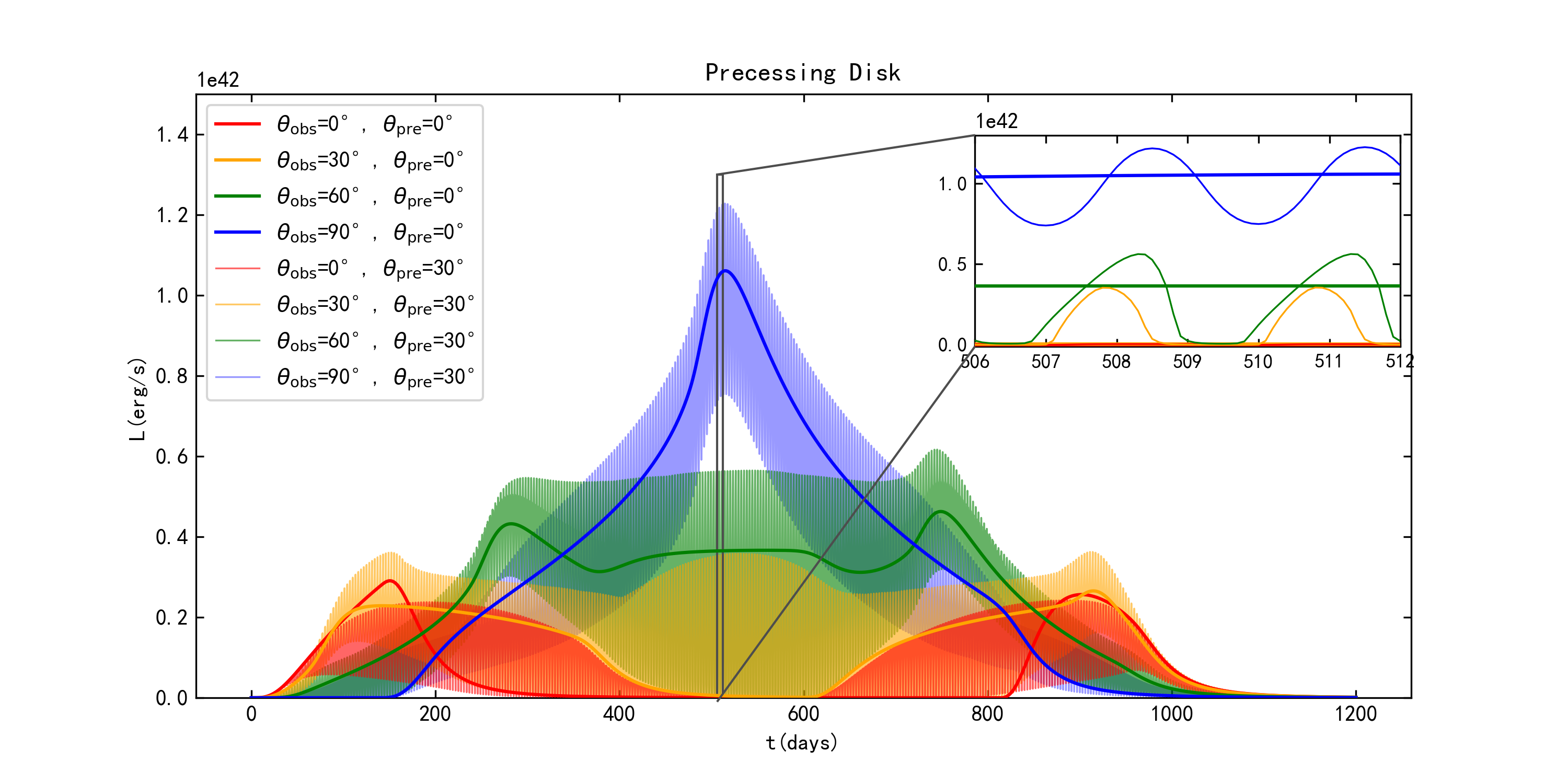}
\caption{The light curve of infrared echoes under the same precession angle and different observation angles. The insert panel in the upper-right corner shows details of the light variation curve from 500 to 506 days. }
\label{fig:01}
\end{figure*}

\begin{figure*}[ht!]
\plotone{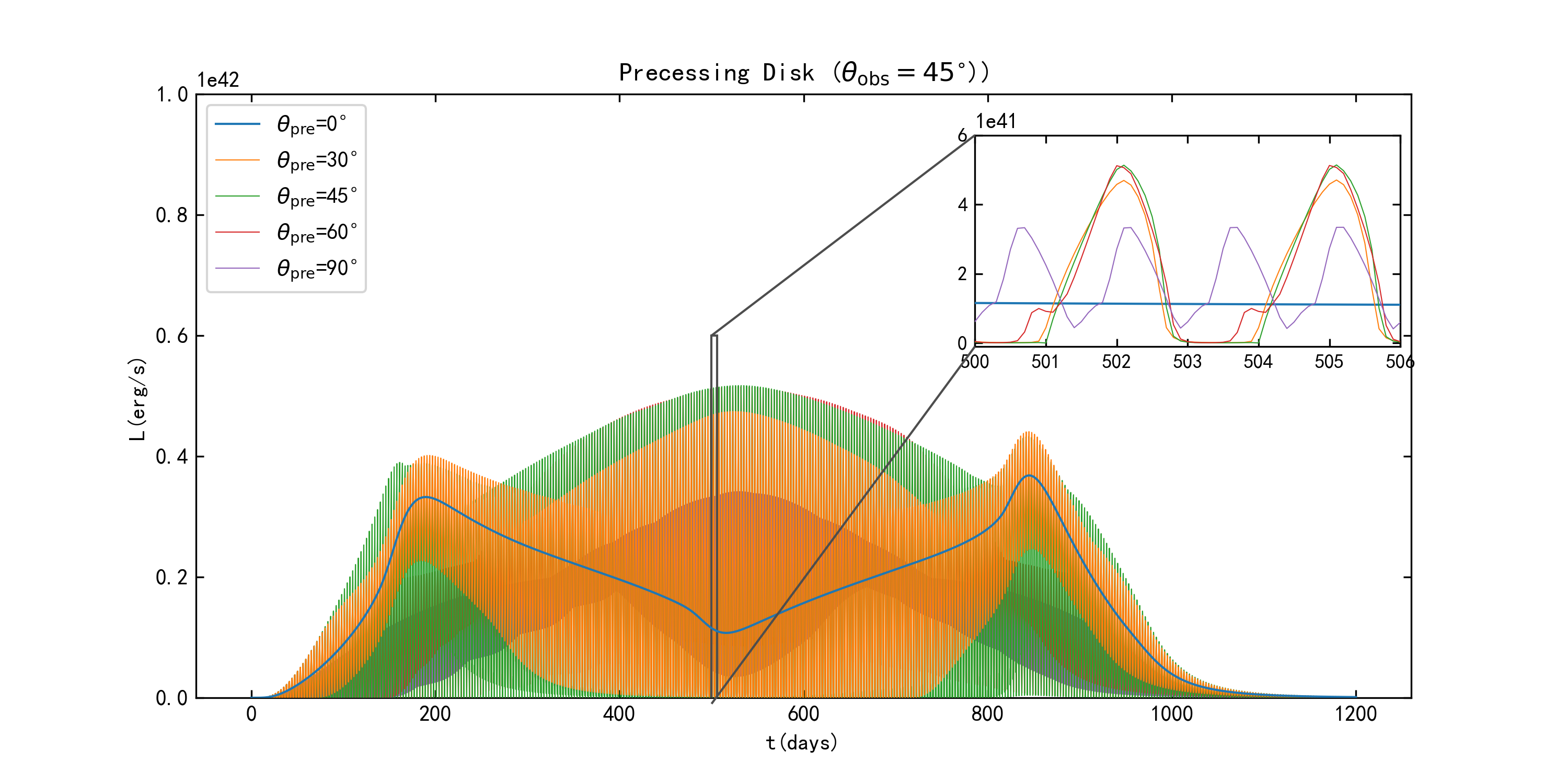}
\caption{Same as Figure \ref{fig:01}, but with a precession angles of $45^{\circ}$. }
\label{fig:02}
\end{figure*}

Using the physical parameters adopted -$L_{disk}=4 \times  10 ^ {45}~\text{erg s}^{-1} $, $\Sigma_{d}=0.1 \text{g cm}^{-2}$, $R=0.4$\, pc, precession period $P_{\rm pre}=3$ days, disk opening angle $45 ^{\circ} $ - we calculated the radiative thermal infrared luminescence curves shown in \autoref{fig:01} and \autoref{fig:02}.

\autoref{fig:01} presents the infrared echo light curves for a precessing disk with fixed precession angle $\theta_{\rm pre} = 30^\circ$, viewed at different observation angles $\theta_{\rm obs}$. The dust ring is inclined at $45^\circ$ relative to the black hole equatorial plane. Combined with the finite ring width and the precession angle, the illuminated region covers only a fraction of the dust ring---making the geometry intermediate between the pure ring and blob models discussed in \autoref{sec:2}. 

At $\theta_{\rm obs} = 0^\circ$, the viewing geometry most closely resembles the pole-on perspective of the blob model, producing a double-peaked light curve as two distinct clumps of illuminated dust arrive at different times due to path-length differences. As $\theta_{\rm obs}$ increases, the path to the near side of the torus becomes longer, delaying the first peak, while the path to the far side becomes shorter, advancing the second peak. And the observer's perspective gradually shifts toward the edge-on configuration, and the light curve morphology transitions from double-peaked to single-peaked structures. However, the precession introduces important modifications to the pure blob picture. Because the illuminated region varies with orbital phase, there are epochs where the geometry temporarily approximates the ring model creating brief plateau-like features superimposed on the otherwise peaked structure. This hybrid behavior is most evident at intermediate viewing angles ($\theta_{\rm obs} \sim 30^\circ$–$60^\circ$), where the light curve shows asymmetric peaks with flattened tops. Additionally, larger observation angles sample dust at progressively higher polar angles $\theta$ on the ring, increasing the characteristic light-travel time across the illuminated region. This explains the systematic delay in peak arrival time from $\sim$180 days at $\theta_{\rm obs} = 30^\circ$ to $\sim$ 500 days at $\theta_{\rm obs} = 90^\circ$, as well as the broader peak profiles at high inclination. 

\autoref{fig:02} isolates the precession-angle effect by fixing $\theta_{\rm obs}=45^\circ$ and varying $\theta_{\rm pre}$. With zero precession ($\theta_{\rm pre}=0^\circ$), the disk's radiation axis remains fixed, illuminating only a single, static patch of the torus. The resulting light curve exhibits two isolated peaks: the first peak corresponds to the near side of the torus, the second to the far side. In this limit, the torus effectively behaves as two disconnected blobs, producing a blob-like signature. 

Introducing precession ($\theta_{\rm pre}>0^\circ$) fundamentally alters this picture. The peak times remain anchored at $\sim 180$ days and $\sim 950$ days, as they are dictated by the light-crossing time to the fixed inner and outer edges of the torus. However, as the disk's radiation cone sweeps across the torus, it progressively heats the entire ring structure between these two extremities. This creates a plateau of re-emission that fills in the previously gap. Critically, the plateau luminosity depends on how closely the precession angle aligns with the torus orientation: $\theta_{\rm pre}$ closer to the torus inclination ($45^\circ$) heats a greater fraction of the ring, yielding a more luminous and broader plateau. Thus, the precession angle does not shift the peak arrival times but controls the duty cycle of heating across the torus: $\theta_{\rm pre}$ aligned with the torus orientation transforms the echo from a double-blob into a continuous ring signature. 

The 3-day precession period is much shorter than the light-crossing time to the dust ($\sim 2R/c \approx 240$ days), causing the heating beam to rapidly sweep the torus many hundreds of times during the echo rise. This rapid scanning imprints a high-frequency modulation on the underlying slow envelope: the IR light curve exhibits fine-scale, 3-day periodic ripples superposed on the broad peak. The inset panels in the upper-right corners of \autoref{fig:output} and \autoref{fig:Model} zoom in on this subtle structure, revealing a train of micro-peaks spaced precisely at the precession period. These features constitute a smoking-gun signature of precession that is independent of the slower geometric delays: if the precession period were longer (e.g., comparable to $2R/c$), the light curve would instead show a single, smooth sweep without substructure. 

\section{Discussion and Conclusion}
\label{sec:4}

\begin{figure*}[ht!]
\plotone{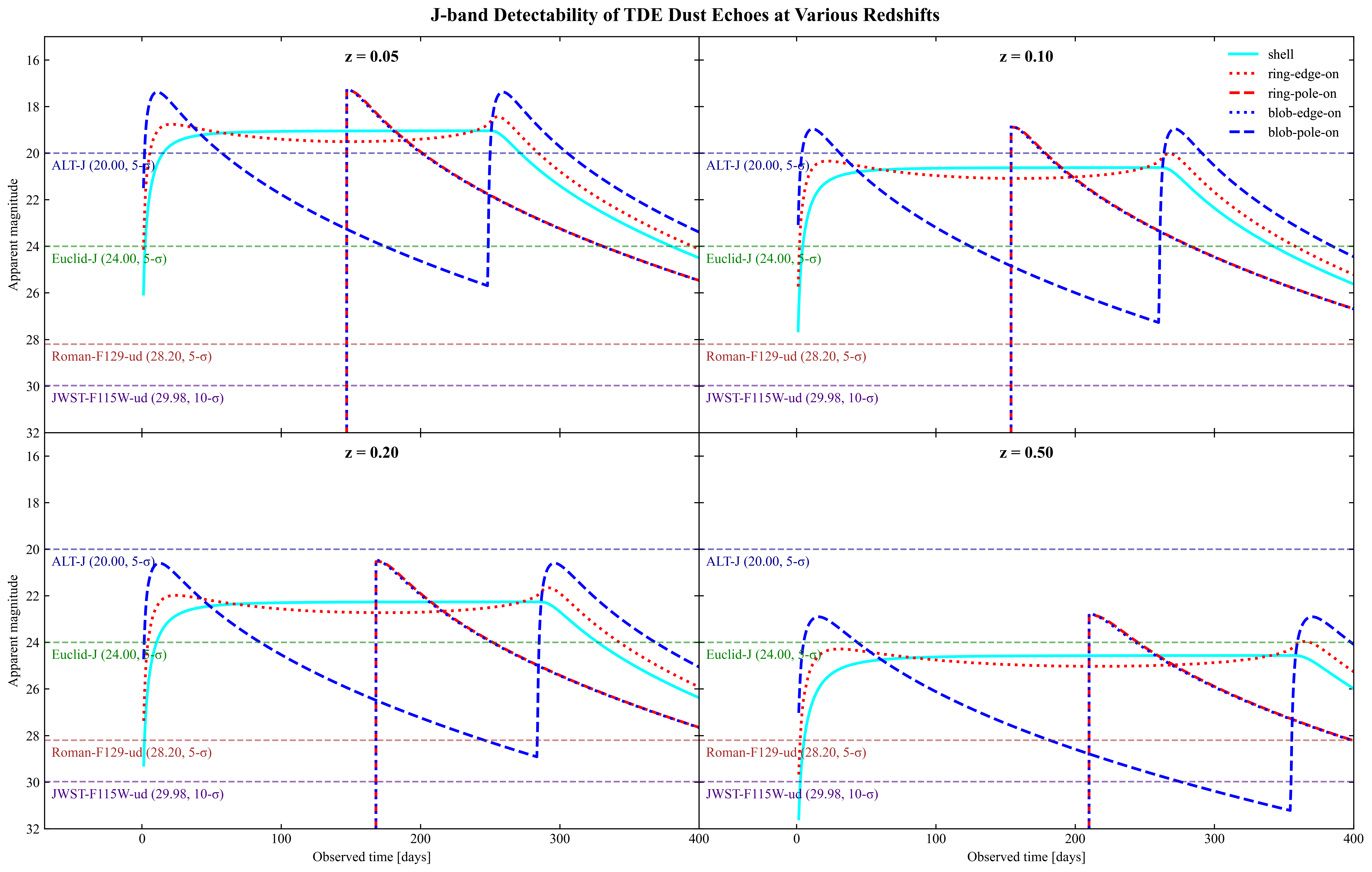}
\caption{Horizontal dashed lines indicate the  point-source limiting magnitudes of near-infrared surveys: Euclid J-band ($ J_E $, 24.0 AB mag, wide survey, 4500 s total exposure, see \citealp{esa_euclid_eoportal}), Roman F129 (J-band equivalent, 28.2 AB mag for Ultra-Deep HLWAS tiers with $\sim 2.4 $ hr total exposure, see \citealp{committee2025romanobservationstimeallocation}), and JWST F115W (29.98 AB mag for JADES Deepest tiers with $\sim 34$ hr total exposure, see \citealp{eisenstein2025overviewjwstadvanceddeep}), and ALT J-band (20.0 AB mag, 4 hr total exposure, see \citealp{2025GCN.42990....1F}).}
\label{fig:z}
\end{figure*}

\subsection{TDE host galaxies and detectability of dust echoes}
\label{sec:4.1}

Tidal disruption events (TDEs) preferentially manifest in compact "green valley" galaxies and post-starburst (E+A) systems, with stellar masses spanning the range of $log (M_{gal}/M_{\odot}) \approx 9.3-11.2$ \citep{Hammerstein_2021, hammerstein2023integralfieldspectroscopy13}. These host galaxies are distinguished by high Sérsic indices (indicating centrally concentrated light profiles), elevated central stellar mass surface densities on scales of $30-100$ pc, and steeper surface brightness profiles compared to typical early-type galaxies \citep{Hammerstein_2021}. The ZTF TDE sample reveals an approximately fivefold overrepresentation of green valley galaxies (comprising $63\%$ of TDE hosts versus $ \sim 13\%$ of SDSS galaxies) and an approximately twenty-twofold overrepresentation of E+A galaxies \citep{hammerstein2023integralfieldspectroscopy13}. The majority of TDE hosts are situated in the green valley, positioned between the star-forming blue cloud and the passive red sequence, exhibiting weak $H_{\alpha}$ emission and pronounced Balmer absorption lines that suggest quiescent or recently terminated star formation \citep{Hammerstein_2021}. 

These host galaxies demonstrate infrared luminosities of $L_{IR} \sim 10^{42}-10^{44}~\text{erg s}^{-1}$, with a median of $\sim 5 \times 10^{43}~\text{erg s}^{-1}$ at $4.6 \mu m$ \citep{Wang_2018}, predominantly attributed to stellar continuum emission in the near-infrared and thermal dust re-radiation in the mid-infrared. The dust echo induced by TDEs contributes a distinct, transient component characterized by peak luminosities of $L_{IR} \sim 10^{42}-10^{43}~\text{erg s}^{-1}$ \citep{Dou_2016} and characteristic dust temperatures of $T \sim 900-2500 K$ \citep{Jiang_2017}. For typical quiescent hosts with $L_{IR, host} \sim 10^{42}-10^{43}~\text{erg s}^{-1}$, the echo signal is comparable to or surpasses the host background, facilitating its straightforward detection through multi-epoch infrared imaging \citep{pasham2025usinginfrareddustechoes}.

However, the detectability degrades significantly for TDEs in star-forming or AGN-dominated hosts with $L_{IR} > 10^{44}~\text{erg s}^{-1}$ \citep{1978ApJ...226..550R}, where the echo represents only a minor perturbation on the total infrared budget. In such environments, sophisticated techniques---including difference images \citep{Jiang_2021, decoursey2025photometricevidencetransientvariablesource}, color-color selection \citep{masterson2024newpopulationmidinfraredselectedtidal, Stern_2012, yao2025distinguishingtidaldisruptionevents}, and spectroscopic decomposition \citep{Liu__2026}---are required to isolate the echo signal from the complex host background. The JWST's unparalleled sensitivity (reaching $L_{IR} \sim 10^{41}~\text{erg s}^{-1}$ at $z \sim 0.5$) will enable dust echo detection even in moderately luminous hosts, transforming TDEs from rare curiosities into systematic probes of circumnuclear dust geometry across diverse galactic environments.

\subsection{Observational Accessibility}
\label{sec:4.2}

\begin{figure*}[htbp]
\gridline{\fig{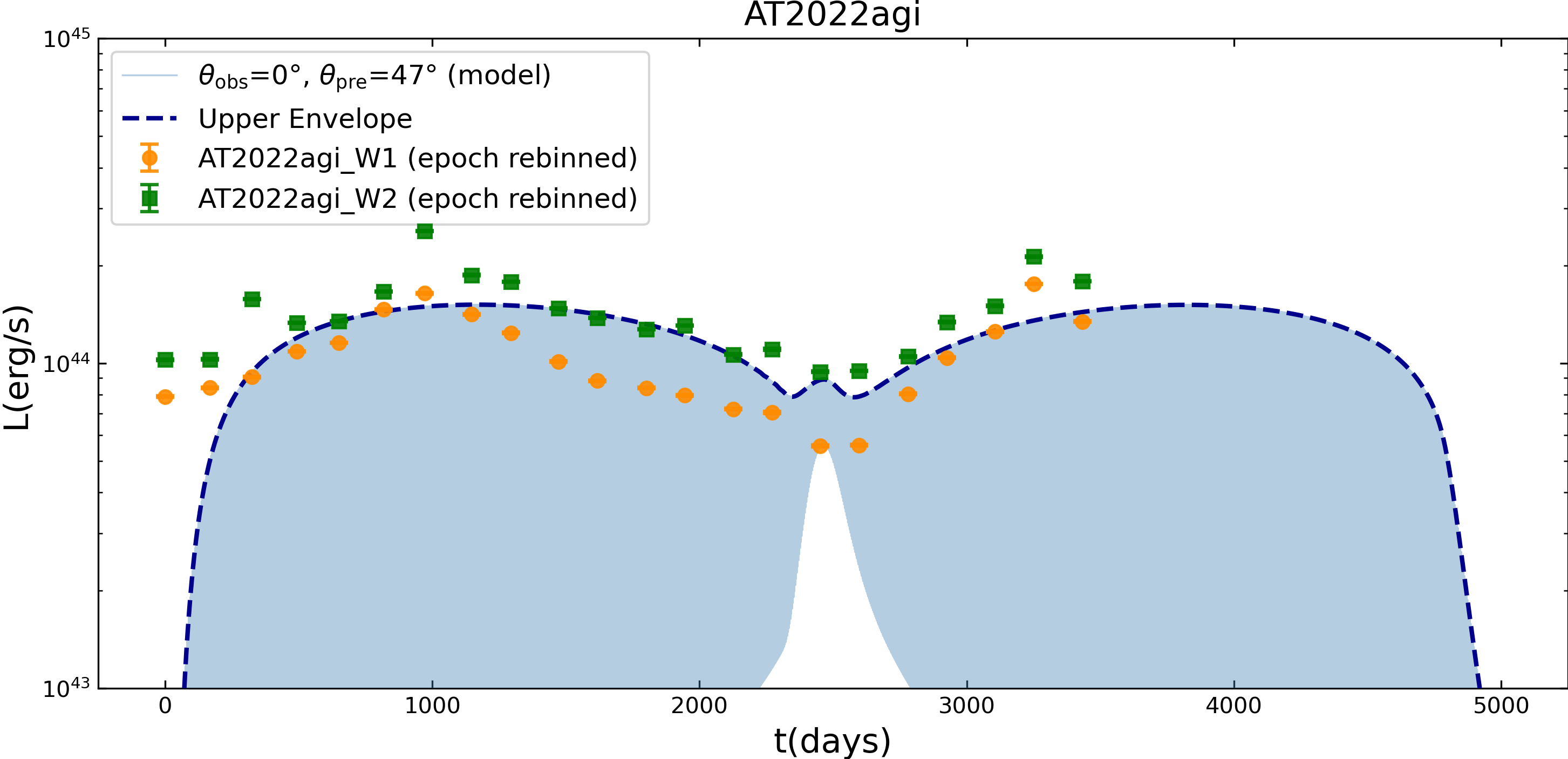}{0.5\textwidth}{(a)}
          \fig{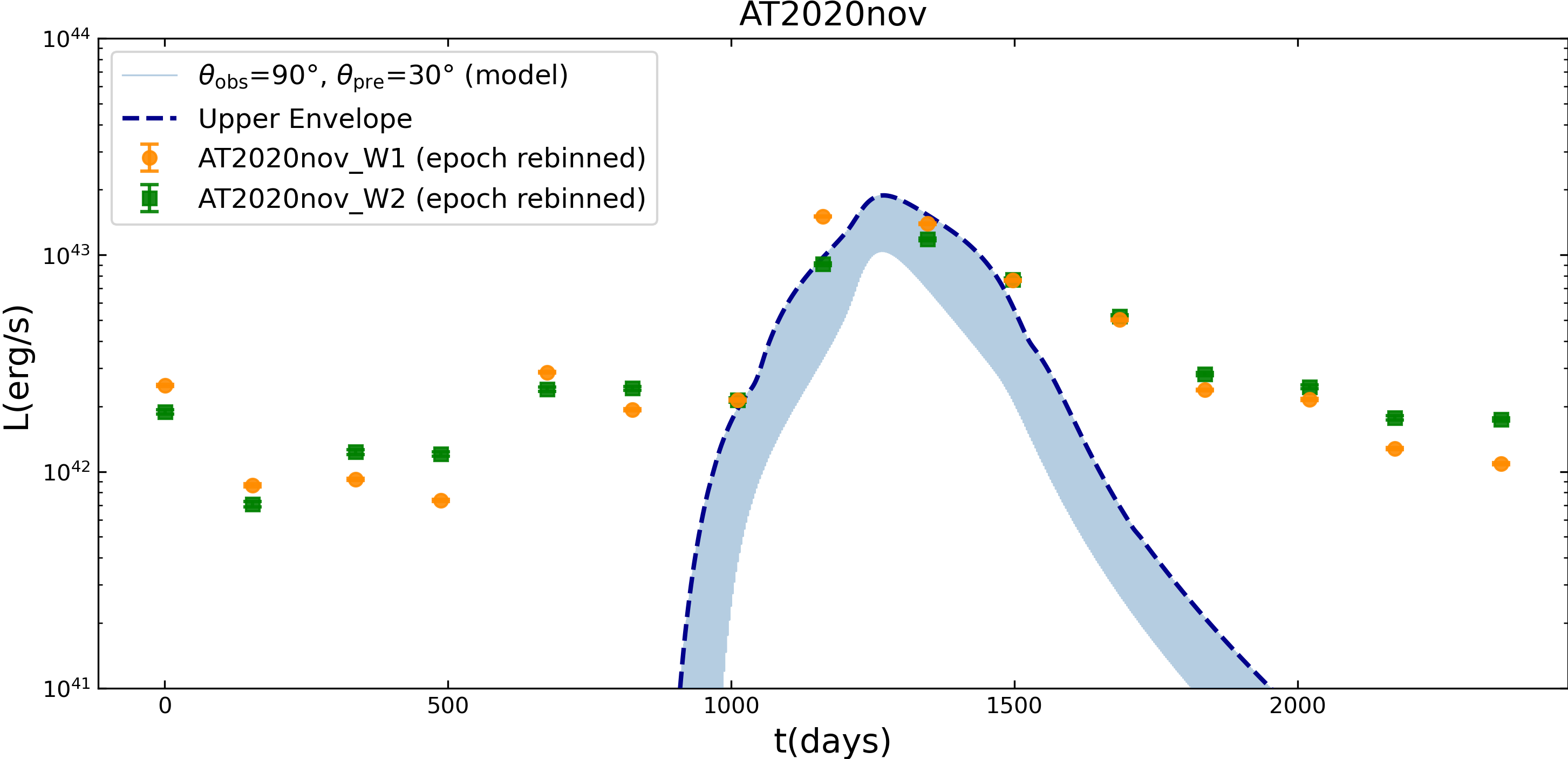}{0.5\textwidth}{(b)}
          }
\caption{Mid-infrared dust echo light curves of (a) AT~2022agi and (b) AT~2020nov. The epoch-binned WISE/NEOWISE W1 ($3.4~\mu$m, orange circles) and W2 ($4.6~\mu$m, green squares) data are shown after host-galaxy baseline subtraction. (a) The model adopts $\theta_{\rm obs}=0^{\circ}$ and $\theta_{\rm pre}=47^{\circ}$ at $R=2.0$~pc, producing a double-humped morphology arising from differential light-travel times to the near-side and far-side dust. (b) The model adopts $\theta_{\rm obs}=90^{\circ}$ and $\theta_{\rm pre}=30^{\circ}$ at $R=0.3$~pc, producing a merged single peak due to the extreme light-travel asymmetry at edge-on viewing geometry. The elevated pre-peak emission at $t \le 1000$ days suggests early dust heating at smaller radii or residual host contamination. The thin solid cyan curves show the theoretical precessing-disk models, and the thick dashed blue curves denote the upper envelopes.}
\label{fig:obs_comparison}
\end{figure*}

%\begin{figure*}[htbp]
%\plotone{AT2022agi_with_envelope_rebinned.png}
%\caption{Mid-infrared dust echo light curves of AT 2022agi in the W1 (3.4 $\mu$m, orange circles) and W2 (4.6 $\mu$m, green squares) bands. The data points represent epoch-binned WISE/NEOWISE photometry after host-galaxy baseline subtraction. The thin solid cyan curve shows the theoretical precessing-disk model with $\theta_{\rm obs}=0^{\circ}$ and $\theta_{\rm pre}=47^{\circ}$ at $R=2.0$ pc. The thick dashed blue curve denotes the upper envelope of the model, which captures the double-humped morphology arising from differential light-travel times to the near-side and far-side dust. }
%\label{fig:AT2022agi}
%\end{figure*}

%\begin{figure*}[ht!]
%\plotone{AT2020nov_with_envelope_rebinned.png}
%\caption{Same as \autoref{fig:AT2022agi}, but for AT 2020nov. The epoch-binned W1 (orange circles) and W2 (green squares) data reveal a single-peaked echo morphology centered near $t \approx 1300$ days, with a peak luminosity of $\sim 10^{43}~\text{erg s}^{-1}$. The model adopts $\theta_{\rm obs}=90^{\circ}$ and $\theta_{\rm pre}=30^{\circ}$ at $R=0.3$ pc, producing a merged single peak due to the extreme light-travel asymmetry at edge-on viewing geometry. The elevated pre-peak emission at $t \le 1000$ days suggests early dust heating at smaller radii or residual host contamination. }
%label{fig:AT2020nov}
%\end{figure*}

In broadband metering, the background equivalent brightness of J is usually lower than H and K, which is beneficial for detecting weak sources. In addition, for small telescopes, the J-band is easier to implement, so the resources in the J-band will be more abundant. \autoref{fig:z} indicates that the detectability of J-band dust echoes in tidal disruption events (TDEs) is influenced by a combination of redshift, dust geometry, and observation depth.Ground-based facilities, exemplified by the Altay Observatory Telescope (ALT), with a limiting magnitude of 20.0 AB mag with 4 hours exposure \citep{2025GCN.42990....1F}, can only detect the earliest peak radiation of recent TDEs within the first 30 days at redshifts $z \lesssim 0.1$. This temporal limitation hinders our ability to fully characterize the evolution of the light curve and identify precession features. 

The wide-field survey observations conducted by the Euclid satellite can achieve a depth of 24.0 AB mag in the J-band \citep{esa_euclid_eoportal}, enabling comprehensive coverage of the light curve evolution for events with redshifts $z \lesssim 0.20$. However, when the redshift reaches $z=0.50$, the echo flux falls below the detection threshold of the Euclid satellite, preventing the facility from observing evolution processes or detecting any periodic substructures imprinted by relativistic precession. 

The Roman Space Telescope has significantly enhanced its observational capabilities through its tiered high-sensitivity wide-field survey (HLWAS) strategy \citep{committee2025romanobservationstimeallocation}. The limiting magnitude of its intermediate observation mode is 26.40 AB mag,The deep and ultra-deep observation modes reach depths of 27.60 and 28.20 AB mag, respectively, fully covering all redshift ranges depicted in the figure. Consequently, the Roman Space Telescope serves as an ideal instrument for constructing a statistical sample of precessing TDEs. Nevertheless, its photometric accuracy and angular resolution may not suffice to detect the most subtle features predicted by rapid precession. 

The F115W band observation depth range of the Near-Infrared Camera (NIRCam) on the James Webb Space Telescope (JWST) spans from 28.87 AB mag (medium depth) to 29.98 AB mag (ultra-deep depth), exhibiting unparalleled sensitivity to faint late-stage echo radiation \citep{eisenstein2025overviewjwstadvanceddeep}. Notably, JWST is the sole observational facility capable of resolving the fine-scale, 3-day periodic ripples predicted by our precession model, which provide conclusive and direction-independent evidence of Lense-Thirring precession. 
%At a redshift of $z=0.50$, the bimodal morphology of bipolar patches observed from the polar axis direction exhibits a peak interval of approximately 150 days, with the second peak magnitude being $m_J \sim 22.5$. This feature remains readily detectable by JWST but poses a considerable challenge for ultra-deep observations with the Roman Space Telescope. 

We advocate implementing a two-tiered observation strategy to maximize scientific returns:

1. Firstly, leveraging the deep and ultra-deep observation modes of the Roman Space Telescope, precessing TDE candidates can be identified based on anomalous light curve shapes, particularly the bimodal structure and asymmetric plateau predicted by deviations from the spherical shell model. 

2. Secondly, JWST's NIRCam Time-Series Imaging enables high-cadence photometric monitoring of confirmed candidates \citep{jwst_nircam_tsi}. Optimized for precise flux measurements of bright sources, this mode supports single uninterrupted epochs of 6–12 hours with minute-level sampling, capturing asymmetric rises, peak profiles, and decay slopes. Phase-diverse coverage is achieved via multiple independent observing blocks strategically distributed across the 2–3 day precession period, accommodating the thermal scheduling constraints that preclude continuous JWST monitoring.

The synergistic integration of Roman's wide-field discovery capabilities with JWST's high-cadence photometric prowess will metamorphose TDE dust echoes from mere passive calorimeters into dynamic tracers of relativistic effects and the geometric configuration of circumnuclear dust. Our model offers explicit, verifiable predictions: precessing TDEs ought to display time-variable plateau luminosities that scale with the precession angle, whereas rapid-cadence JWST photometry has the potential to detect the hourly-to-daily periodic ripples that serve as definitive signatures of Lense-Thirring precession. The detection of these features will facilitate direct measurements of black hole spin and the three-dimensional dust structure, thereby establishing TDEs as precision laboratories for relativistic astrophysics in the era of JWST. 

\subsection{Application to Observed TDEs}
\label{sec:4.3}

To illustrate the observational significance of our precessing TDE dust-echo model, we apply it to two extensively studied tidal disruption events with comprehensive WISE mid-infrared monitoring: AT 2020nov and AT 2022agi. Both sources exhibit prolonged mid-infrared emission characteristic of dust echoes, providing empirical testbeds for constraining the geometric parameters of our framework.

\subsubsection{Data Reduction and Epoch Binning}
\label{sec:4.3.1}

The infrared data for both sources are retrieved from the TDEcat catalog \citep{Langis_2026}, which compiles WISE and NEOWISE multiepoch photometry in the W1 (3.4 $\mu$m) and W2 (4.6 $\mu$m) bands. The nominal observational cadence is approximately six months, with each epoch comprising multiple individual exposures acquired during a single WISE visit ($\sim 1$ day duration). 
We process the raw photometry through two steps. First, we subtract the host-galaxy baseline emission to isolate the TDE-related dust echo. For AT 2022agi, the baseline magnitudes are adopted as $m_{W1} = 12.88$  and $m_{W2} = 11.26$ \citep{Dou_2017}, corresponding to the quiescent infrared luminosity of the host. For AT 2020nov, we compute the baseline from the average luminosity of all pre-outburst epochs observed prior to MJD 58000.0, ensuring that the reference level captures the host contribution without contamination from the rising TDE flare. The baseline-subtracted luminosities are then converted to absolute units for direct comparison with our model predictions. 
Second, we rebin the individual exposures into epoch-level measurements to enhance the signal-to-noise ratio. Within each WISE visit, exposures are separated by less than one day, while successive visits are spaced by $\sim 180$ days. We group exposures using a 10-day time-separation threshold, which effectively merges all frames from a single visit into one representative data point. For each epoch, we adopt the median observation time and the mean luminosity, with uncertainties propagated as the standard error of the mean. This binning strategy preserves the temporal resolution necessary to trace the broad echo morphology while suppressing short-timescale photometric scatter. 

\begin{deluxetable}{lcc}[ht!]
\tablecaption{Best-fitting Precessing TDE Model Parameters for AT~2022agi and AT~2020nov}
\label{tab:model_params}
\tablehead{
\colhead{Parameter} & \colhead{AT 2022agi} & \colhead{AT 2020nov}
}
\startdata
Dust torus radius $R$ (pc) & 2.0 & 0.3 \\
Viewing angle $\theta_{\rm obs}$ ($^\circ$) & $0$ & $90$ \\
Precession angle $\theta_{\rm pre}$ ($^\circ$) & $47$ & $30$ \\
Peak luminosity (erg s$^{-1}$) & $\sim 7 \times 10^{46}$ & $\sim 8 \times 10^{45}$ \\
Echo morphology & Double-peaked & Single-peaked \\
\enddata
\tablecomments{The viewing angle $\theta_{\rm obs}$ is measured between the observer's line of sight and the black hole spin axis (see Section \ref{sec:3}). The precession angle $\theta_{\rm pre}$ denotes the tilt of the accretion disk spin axis relative to the black hole spin axis. The peak luminosity represents the total TDE peak luminosity, i.e., the input energy for the infrared echo. Both sources are fitted with a fixed  precession period $P_{\rm pre} = 3$ days. }
\end{deluxetable}

\subsubsection{AT 2022agi}
\label{sec:4.3.2}

AT 2022agi is a repeating tidal disruption event in the ultra-luminous infrared galaxy F01004-2237 (z=0.11783). The first optical flare was detected in 2010 by the Catalina Real-Time Transient Survey  \citep{tadhunter2017tidaldisruptioneventnearby}, while a second flare erupted in 2021 and was independently discovered by ATLAS and Gaia \citep{Sun_2024}. The two flares are separated by $10.3 \pm 0.3$  years, with the second flare reaching a peak luminosity of $4.4 \times 10^{44}~\text{erg s}^{-1}$ and exhibiting very broad emission lines characteristic of TDEs. We analyze the mid-infrared dust echo following the 2010 flare using multiepoch WISE/NEOWISE photometry spanning approximately 5000 rest-frame days. \autoref{fig:obs_comparison} (a) displays the W1 and W2 epoch-binned luminosities alongside our model predictions. The data reveal a double-humped structure across the entire observational baseline, with peak luminosities reaching approximately $ 2 \times 10^{44} ~\text{erg s}^{-1} $ in both bands. 

We determine the best-fitting model parameters by comparing with the upper envelope of the precessing light curves. The observed morphology, characterized by two distinct peaks separated by approximately 2500 days and a pronounced central dip near $t \approx 2500$  days, resembles the theoretical signature of a precessing disk viewed near edge-on. Specifically, the double-peaked structure arises from the differential light-travel times between the near-side and far-side dust illumination. 

Our model favors a precession angle $\theta_{\rm pre} \approx 47^{\circ}$ and a viewing angle $\theta_{\rm obs} \approx 0^{\circ}$ (near edge-on , the angle is $45^{\circ}$ between the ring and the disk), with a dust torus radius $R = 2.0$  pc. The near edge-on viewing geometry naturally produces the symmetric double-peak structure, while the substantial precession angle---closely aligned with the torus inclination---ensures that the heating duty cycle spans a significant fraction of the torus, filling in the inter-peak region with a luminous plateau. The model upper envelope successfully captures the peak positions and the broad echo extent, although the observed peak-to-plateau contrast is somewhat less pronounced than predicted, possibly indicating a more extended dust distribution or non-uniform surface density within the torus. 

\subsubsection{AT 2020nov}
\label{sec:4.3.3}

AT 2020nov presents a contrasting case with a more compact echo evolution, as illustrated in \autoref{fig:obs_comparison} (b). The WISE monitoring covers approximately 2000 rest-frame days, during which the mid-infrared emission exhibits a single dominant peak centered near $t \approx 1200$  days, reaching approximately $10^{43} ~\text{erg s}^{-1}$. Notably, the pre-peak baseline (at $t < 1000$ days) shows elevated emission at approximately $10^{42}~\text{erg s}^{-1}$, significantly above typical quiescent host levels, suggesting either early dust heating from the initial TDE flare at smaller radii or residual host contamination imperfectly subtracted by the baseline. 

The single-peaked morphology is consistent with our model predictions for viewing geometries approaching pole-on. We find optimal agreement with a viewing angle $\theta_{\rm obs} \approx 90^{\circ}$ and a precession angle $\theta_{\rm pre} \approx 30^{\circ}$. At this near-pole-on inclination, the light-travel path to the near side of the torus is maximally lengthened while the far-side path is shortened, causing the two geometric peaks to merge into a single broad maximum. The pre-peak baseline emission likely originates from the early illumination of the near-side dust before the main peak arrives, a feature that our simplified thin-shell model with uniform surface density captures only qualitatively. 

The lower overall luminosity of AT 2020nov relative to AT 2022agi (approximately $10^{43}$ versus $ 2 \times 10^{44}~\text{erg s}^{-1}$) may reflect a lower peak bolometric luminosity of the central TDE flare, a smaller dust covering factor, or a more distant dust distribution. The model normalization, scaled by the dust mass and surface density, provides an approximate match but suggests that a more detailed treatment of the radial dust profile would be required for precise calorimetric constraints. 

\subsubsection{ Implications for Precession Diagnostics}
\label{sec:4.3.4}

The contrasting morphologies of AT 2022agi (double-peaked) and AT 2020nov (single-peaked) illustrate how the viewing geometry fundamentally shapes the observable dust-echo signature within a unified precessing-disk framework. Both sources are consistent with the same underlying physical model, with best-fitting parameters summarized in Table \ref{tab:model_params}.

Notably, neither source exhibits the fine-scale periodic ripples predicted for short precession periods (e.g., $P_{\rm pre}=3$  days; see  Section \ref{sec:3}). This absence is physically reasonable: the WISE sampling cadence (approximately 180 days) is far too coarse to resolve sub-day or multi-day precession modulations, and the epoch-binning procedure further suppresses such high-frequency variability. Detection of these ripples would require high-cadence monitoring with facilities such as JWST or the Roman Space Telescope, as discussed in Section \ref{sec:4.2}. 

The qualitative reproduction of the broad echo envelopes for both sources validates the precessing TDE model as a viable framework for interpreting mid-infrared dust echoes. Future work incorporating multi-band spectral energy distributions, spatially resolved dust geometries, and more sophisticated radial dust profiles will enable tighter constraints on the precession parameters and the circumnuclear dust properties.

\begin{figure*}[htbp]
\plotone{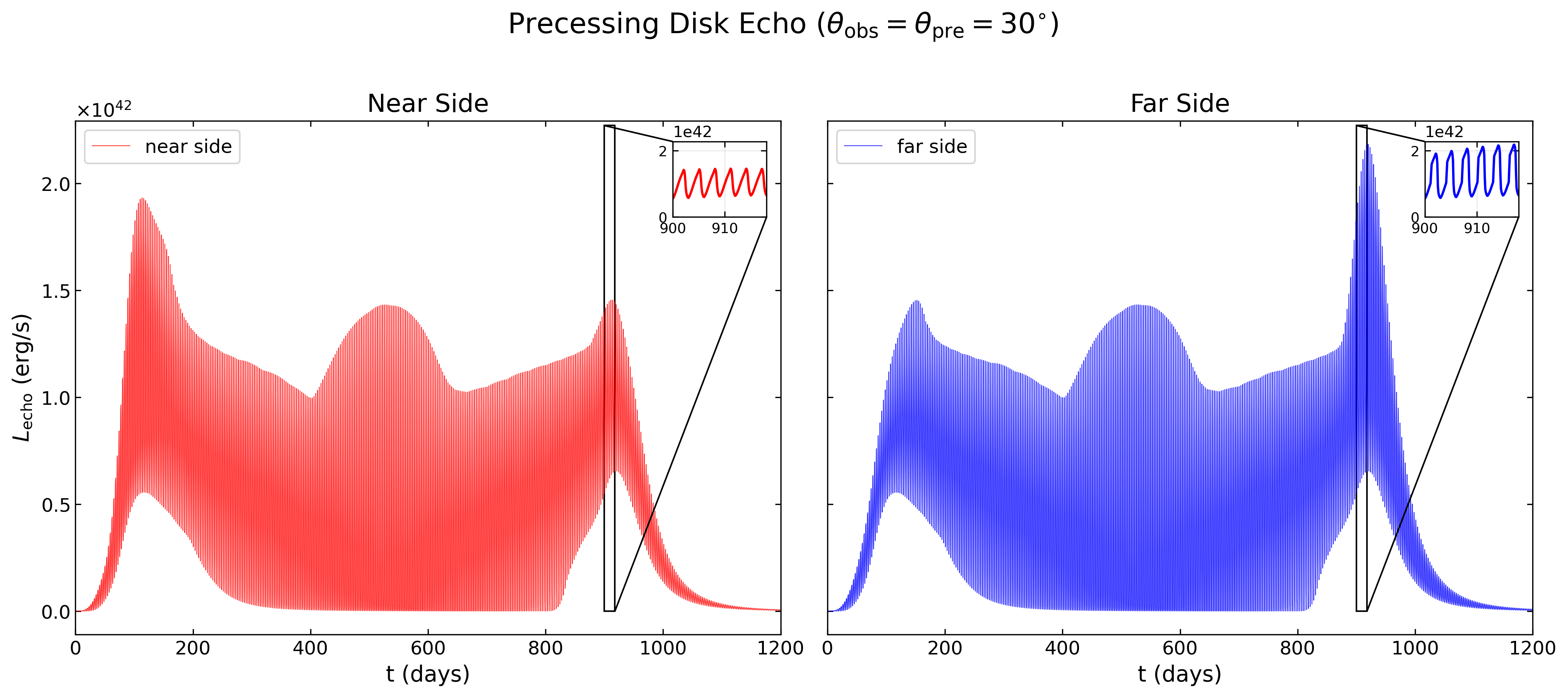}
\caption{Dust echo light curves for a precessing disk ($\theta_{\rm obs} = \theta_{\rm pre} = 30^{\circ}$) with a fixed high-density clump ($10\times$ ambient density, $10^{\circ}$ angular size) placed at different angular positions. The near-side (red) and far-side (blue) responses show identical precession-modulated envelopes, but the local peak luminosity is significantly enhanced at the phase where the precessing beam sweeps across the clump. The inset panel in the upper-right corner shows the detailed light variation from 900 to 918 days, revealing the rapid periodic modulation from disk precession.}
\label{fig:echo_comparison}
\end{figure*}

\subsection{Model Limitations}
\label{sec:4.4}

Our model adopts several simplifications that merit discussion. First, we assume a single grain size and uniform dust composition. Real circumnuclear dust features a continuous size distribution, with smaller grains sublimating at larger radii than larger ones, which smooths the sharp inner boundary of our thin-shell model. Yet recent modeling of the TDE AT2019qiz demonstrates that the inferred torus geometry---inclination, opening angle, and inner radius---is nearly independent of the assumed dust grain size and composition; only the surface density (which scales the overall luminosity) is sensitive to these details \citep{Wu_2025}. This indicates that the large-scale echo morphology, including the rise, plateau, and double-to-single peak transition, is governed by geometry rather than grain microphysics.

Second, to test the effect of spatial non-uniformity, we used densities with random fluctuations at different $\phi$ and $\theta$ positions for calculation. The resulting light curves retain the same overall shape, with only the peak luminosity changing, because the echo luminosity integrates over the full radiating surface and averages out small-scale density variations. Although this addresses continuous inhomogeneity, we note that a clumpy medium \citep{Nenkova_2002,Nenkova_2008_1,Nenkova_2008_2} could behave differently: as the precessing beam sweeps across discrete clumps, the echo luminosity at the corresponding phase may be significantly enhanced while the precession periodicity remains unchanged. To test this scenario, we adopt a two-phase dust model consisting of a smooth background medium with embedded discrete clumps. The dynamical timescale of dust clumps at parsec-scale distances from the black hole ($\sim 10^3$--$10^4$ yr) is much longer than the infrared echo timescale ($\sim 10^2$--$10^3$ d), so the clump positions can be regarded as stationary during the echo evolution \citep{Lyu__2019}. We place a high-density clump (10$\times$ ambient density, angular size $10^{\circ}$) at a fixed angular position and compute the echo light curve for a precessing disk with $
\theta_{\rm obs} = \theta_{\rm pre} = 30^{\circ}$. When the precessing beam sweeps across the clump, the local peak luminosity at that phase is significantly enhanced compared to the smooth envelope, while the  periodicity of the light curve remain unchanged as \autoref{fig:echo_comparison}. This demonstrates that in a quasi-static clumpy medium, discrete clumps amplify the echo luminosity at specific orbital phases without introducing additional temporal structure or altering the underlying precession periodicity.

Third, our thin-shell approximation neglects the radial extent of the dust. Any spherically symmetric distribution can be described as a convolution of shells \citep{maeda2015constrainingcircumstellarmatterdust}, and multi-shell distributions would superpose echoes from different radii, smoothing sharp features. This likely explains why our AT2020nov model in \autoref{fig:obs_comparison} (b) produces a narrower peak than observed: the observed pre-peak baseline emission and the broad peak profile suggest contributions from an extended radial distribution or an inner dust component at smaller radii, which our simplified single-shell model captures only qualitatively. 

Fourth, the assumption of instantaneous thermal equilibrium is justified because the radiative cooling time of dust grains is negligible on all relevant timescales \citep{Gerardy_2002}. Even for the 3-day precession period, dust temperatures track the illumination beam faithfully.

Finally, a realistic grain size distribution smooths the inner sublimation boundary over a finite radial extent, depending on the maximum grain size and the luminosity gradient across the boundary \citep{Nenkova_2008_2,Reynolds_2026}. This radial smoothing, combined with a continuous distribution of dust beyond the inner edge, superposes echoes from multiple radii with slightly different light-travel delays. The net effect is a phase mixing that damps the amplitude of the fine-scale 3-day precession ripples by a factor of a few, but does not erase their periodicity. We estimate that the residual ripple amplitude is damped to the order of a few percent relative to the echo peak---still well within the reach of JWST NIRCam time-series imaging \citep{jwst_nircam_tsi}. The slower large-scale features---the double-to-single peak transition, the asymmetric plateau, and the peak arrival times---are geometric in origin and remain statistically robust against dust microphysics variations \citep{Wu_2025}. Thus, precession-induced signatures remain detectable even under realistic dust conditions, provided that high-cadence observations are obtained.

\subsection{Distinguishing Precession Echoes from Early-Time Hydrodynamic Processes}
\label{sec:4.5}

In realistic TDEs, early-time hydrodynamic processes---stream-stream collisions and chaotic circularization of stellar debris---release significant UV/optical energy prior to the formation of a steady disk \citep{Jiang__2016,steinberg2023streamdiskshocksoriginspeak,Andalman_2021}. Because these processes irradiate the same parsec-scale circumnuclear dust as the later accretion disk, they could, in principle, contribute to the observed IR echo. However, the temporal evolution of the echo naturally separates these components. 

Stream-stream collisions and circularization shocks are transient events operating on dynamical timescales of days to weeks, with their associated UV/optical emission fading on timescales of months \citep{Gezari_2017,huang2025xrayvariabilityphotosphereevolution}. The IR dust echo responds to the incident radiation history through a convolution with the geometric delay kernel \citep{Dwek1983,Dwek1985,maeda2015constrainingcircumstellarmatterdust}. Consequently, a transient early flare produces only a transient early echo component that decays within the light-crossing time of the dust distribution ($\sim 2R/c$). By contrast, the precessing accretion disk provides sustained illumination over years, generating a long-lived echo that dominates the late-time IR light curve. 

Observationally, ASASSN-15oi provides a clear template: the initial UV/optical peak attributed to stream circularization faded within months, while delayed X-ray emission from the newly formed disk appeared only $\sim$ 1  year later \citep{Gezari_2017}. By analogy, the late-time IR echo ($t \gtrsim 2R/c $) traces the sustained, anisotropic illumination of the precessing disk, long after the transient hydrodynamic flare has decayed. Any residual contamination from early shocks would appear as a minor, rapidly fading precursor superposed on the initial rise, not as a periodic modulation.

Crucially, the periodicity of the precession-induced ripples (3-day modulations; \autoref{sec:3}) provides an unambiguous discriminator. Stream-stream collisions and circularization shocks are stochastic, single-impulse events with no intrinsic periodicity. Their dust echoes would exhibit smooth, monotonic rises and declines, whereas precession imprints strictly periodic signatures that persist for the entire echo duration. High-cadence JWST photometry can therefore distinguish these mechanisms even if their early-time contributions partially overlap.

{\section{Summary}}
\label{sec:5}

In this work, we have presented a comprehensive theoretical framework for calculating infrared echoes from precessing tidal disruption events, exploring how the slow slewing of a misaligned accretion disk imprints distinctive signatures on the temporal morphology of dust re-emission. 

Through the calculation of the model we provided, we have obtained the following main conclusions:

1. In the precessing model, the observation angle $\theta_{\rm obs}$ controls the relative timing and morphology of the echo structure. As $\theta_{\rm obs}$ increases, the path to the near side of the torus becomes longer, delaying the first peak, while the path to the far side becomes shorter, advancing the second peak. This progressively compresses the time separation between the two peaks, and at sufficiently large viewing angles ($\theta_{\rm obs} \gtrsim 90^\circ$), the peaks merge into a single-peaked structure characteristic of the edge-on ring geometry. The precession introduces important modifications to this pure blob picture. Because the illuminated region varies with orbital phase, there are epochs where the geometry temporarily approximates the ring model, creating brief plateau-like features superimposed on the otherwise peaked structure. This hybrid behavior is most evident at intermediate viewing angles ($\theta_{\rm obs} \sim 30^\circ - 60^\circ $), where the light curve shows asymmetric peaks with flattened tops.

2. The precession angle does not shift the peak arrival times---these remain fixed by the torus geometry and light-crossing time. Instead, $\theta_{\rm pre}$ controls the heating duty cycle across the ring: precession angles aligned with the torus inclination (e.g., $\theta_{\rm pre} \approx 45^\circ$ for our adopted geometry) sweep the greatest fraction of the torus, producing the most luminous and broadest plateau. This transforms the echo from isolated double-blob signatures (at $\theta_{\rm pre}=0^\circ$) into continuous ring-like emission, with the plateau luminosity peaking when $\theta_{\rm pre}$ matches the torus opening angle. 

3. The 3-day precession period much shorter than the light-crossing time causes the radiation beam to rapidly sweep the torus hundreds of times during the echo rise, imprinting fine-scale 2-day periodic ripples on the broad light curve

These results establish that IR echoes are not merely passive reprocessors but dynamic tracers of the central engine's evolving illumination pattern, offering new avenues to constrain dust covering factors, torus geometries, and relativistic precession mechanisms in galactic nuclei. 

%%The advent of JWST has revolutionized the study of TDE IR echoes, enabling detailed spectroscopic characterization previously unattainable. Recent JWST observations of IR-selected TDEs have revealed compact, accretion-driven emission lines and exceptionally strong silicate emission features, exceeding those seen in typical AGN \citep{masterson2025jwstsviewtidaldisruption}. These discoveries provide unprecedented constraints on the dust mineralogy, optical depth, and the time-dependent radiative transfer within the dusty torus. For instance, modeling of mid-IR flares using time-dependent equal-delay surfaces has allowed direct measurement of torus covering factors ($\sim10\%$) and inner dust radii, linking the IR echo to the immediate post-disruption accretion state \citep{Jiang_2017}. The angular resolution of JWST also enables spatially resolved studies of the dust distribution in nearby events, moving beyond pure photometric constraints. 

%% For this sample we use BibTeX plus aasjournalv7. bst to generate the
%% the bibliography. The sample7. bib file was populated from ADS. To
%% pdflatex sample7. tex
%% pdflatex sample7. tex
%% pdflatex sample7. tex

\begin{acknowledgements}
We thank the participants of the TDE FORUM (Full-process Orbital to Radiative Unified Modeling) online seminar series for their inspiring discussions. We are very grateful to Yanan Wang, Rongfeng Shen, Erlin Qiao, and Chichuan Jin for their helpful discussions. This work is supported by the National Key R\&D Program of China (No. 2023YFC2205901) and the National Natural Science Foundation of China under grants 12473012 and 12533005, and the Fundamental Research Funds for the Central Universities, HUST (No. YCJJ20252115). W. H. Lei. acknowledges support by the science research grants from the China Manned Space Project with NO. CMS-CSST-2021-B11. 
\end{acknowledgements}

\bibliography{Ref}{}
\bibliographystyle{aasjournalv7}

%% the collaboration and author truncation commands are used. It has to


\end{document}